\documentclass[twoside]{article}

\usepackage{lipsum} 

\usepackage[sc]{mathpazo} 
\usepackage[T1]{fontenc} 
\usepackage{microtype} 
\usepackage{wrapfig}

\usepackage{amsmath}

\usepackage{mathtools}
\usepackage[hmarginratio=1:1,top=32mm,columnsep=20pt]{geometry} 
\usepackage{multicol} 
\usepackage[hang, small,labelfont=bf,up,textfont=it,up]{caption} 
\usepackage{booktabs} 
\usepackage{float} 

\usepackage{lettrine} 
\usepackage{paralist} 

\usepackage{abstract} 

\usepackage{titlesec} 
\renewcommand\thesection{\Roman{section}} 
\renewcommand\thesubsection{\Roman{subsection}} 
\titleformat{\section}[block]{\large\scshape\centering}{\thesection.}{1em}{} 
\titleformat{\subsection}[block]{\large}{\thesubsection.}{1em}{} 

\usepackage{fancyhdr} 
\title{\vspace{-15mm}\fontsize{24pt}{10pt}\selectfont\textbf{Periodic almost-Schr\"{o}dinger equation for quasicrystals}} 

\author{
\large
\textsc{Igor V. Blinov}\\[2mm] 
\normalsize Moscow Institute of Physics and Technology\\ 
\normalsize blinov@phystech.edu 
\vspace{-5mm}
}
\date{}

\begin{document}

\maketitle 

\thispagestyle{fancy} 


\begin{abstract}

\noindent A new method for finding electronic structure and wavefunctions of electrons in quasiperiodic potential is introduced. To obtain results it  uses slightly modified Schr\"{o}dinger equation in spaces of dimensionality higher than physical space. It enables to get exact results for quasicrystals without expensive non-exact calculations. New method tested on one-dimensional quasiperiodic screened Coulomb potential.

\end{abstract}


\begin{multicols}{2} 

\section{Introduction}

\lettrine[nindent=0em,lines=3]{Q}uasicrystals after Dan Shechtman's \cite{dan}  discovery became that missing link between periodic crystals and amorphous solids. The new state of matter was in focus of particular interest because of the interesting electronic \cite{poon}, magnetic \cite{kraus}, and elastic properties. Just as periodic crystals, quasicrystals have discrete diffration pattern \cite{dan}, though it combined with the lack of periodicity \cite{dan}, \cite{steinhardt}, which allows quasicrytals to have forbidden for periodic crystals symmetries. 

An obvious idea for describing quasiperiodic crystals is to use so called crystalline approximants - periodic structure with a very large unit cell which emulates a quasiperiodic structure.

Another idea was developed in work of KKL \cite{kitaev} where they used normalization method for a one-dimensional crystal and obtained first exact result for one-dimensional quasiperiodic structures. Unfortunately, there are no exact results for higher dimensional quasicrystals.

Another obvious idea for solving the electronic structure of quasicrystals is to use quasiperiodic functions. Or, more precisely, their periodic images embedded in higher-dimensional spaces.

This idea follows from the fact that qusiperiodic functions satisfy the desired properties: they are non-periodic and their diffraction pattern is discrete \cite{levitan}.
\\

This approach has advantages such as follows: 
\begin{enumerate} 
\item both numerical and analytical calculations are almost as easy as for periodic structures
\item errors of numerical calculations are related only to machine precision
\end{enumerate}

Nevertheless, there is a problem on this way waiting for us: how to embed our physical equations in this non-physical space. \\
In this work I have tried to find a satisfying way to overcome this problem.
\section{Quasiperiodic functions}
Continuous quasiperiodic functions, as a subset of continuous almost periodic functions might be presented in Fourier series:
\begin{equation}\label{f-sum}
f(x)=\sum A_{\kappa} e^{i\Lambda_{\kappa} x}
\end{equation}
where $\Lambda_{\kappa}$, called Fourier indices are real. Unlike periodic functions, where these numbers are integers, in almost periodic functions they might be even dense. And 
\begin{multline}\label{scalar-multiply}
A_{\kappa}=(f(x), e^{i\Lambda_{\kappa} x})=\\=\frac{1}{\int_{-\infty}^{+\infty}dx}\int_{-\infty}^{+\infty}dx f\cdot e^{-i\Lambda_{\kappa} x} .
\end{multline}
\begin{center}
\centering\includegraphics[width=2in]{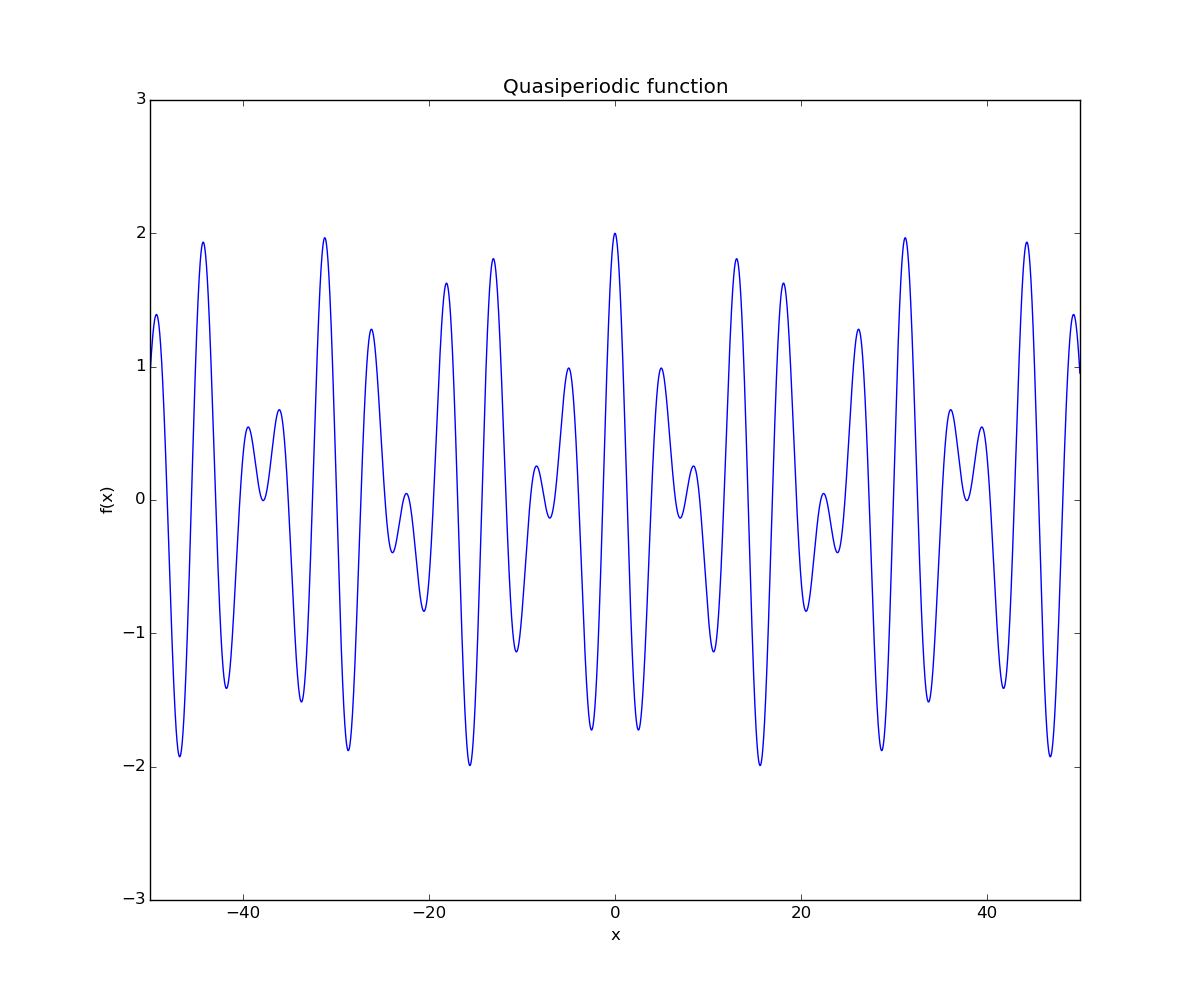}\\
\small \textit{ Fig.1 Quasiperiodic function $f(x)=sin(x)+cos(\sqrt{2}x)$ }
\label{pic_periodic_cos}
\end{center}
It is widely known, that these functions are images of periodic functions on the space of higher dimension.
For example, we may take simple quasiperiodic 1-dimensional function with two quasiperiods (Fig.1)
\begin{equation} \label{cosx}
f(x)=sin(x)+cos(\sqrt{2}x)
\end{equation}
\begin{center}
\centering\includegraphics[width=2in]{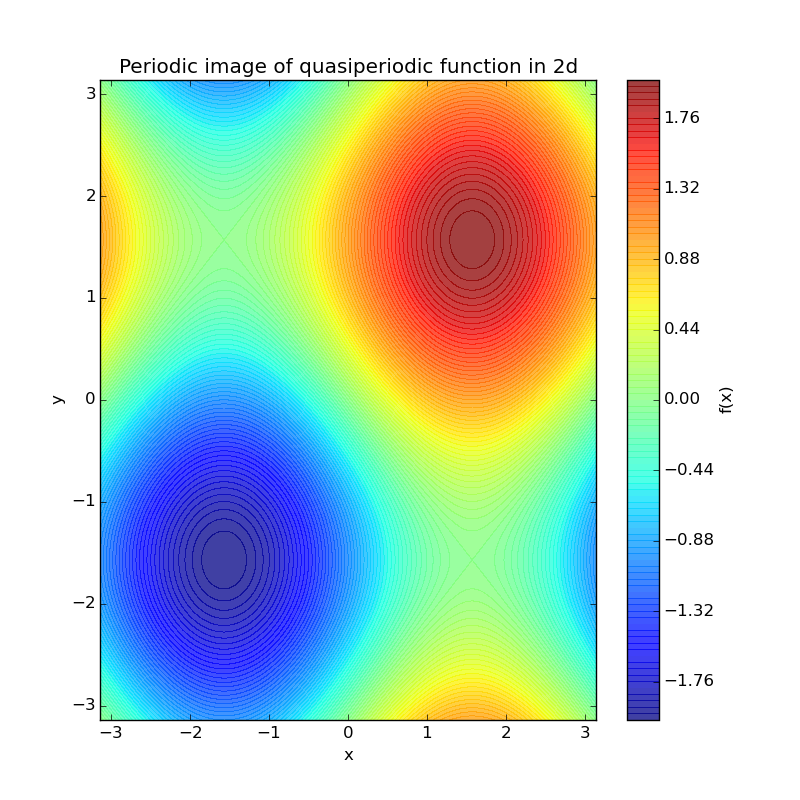}\\
\small \textit{ Fig.2 Periodic image of $f(x)=sin(x)+cos(\sqrt{2}x)$ }
\label{pic_periodic_cos}
\end{center}
and, with transformation $y=\sqrt{2}x$ perform it as periodic in 2-dimensional space (Fig.2).


\section{Physical Intuition}
As it was said previously, we may embed a quasiperiodic one-dimensional function with two incompensurated periods in a two-dimensional space with transformation $y=\tau x$, where $\tau$ is a ratio of quasiperiods.
So, our transformation (x $\times$ y) $\rightarrow$ x looks like moving along the line $y=\tau x$ and "measuring" value of our function. If we will try to imagine, how it looks in a unit cell, this should be moving along family of lines $y=\tau x + \eta$. The exact position on fixed line will be given by $\xi=x+\frac{y}{\tau}$.\\

The same "measuring" of a value of potential is what actually electron do. Embedded in a unit cell $[-a,a]\times [-b,b]$, its movement may look like going from (0,0) along the line $y=\tau x$, then, on the top border $y=b$ (if $\tau >1$), it will jump to the bottom border without changing x. After that, it will start going by another line and so on.\\

So, it might be rational to look for an equation connecting a two-dimensional image of one-dimensional wavefunction and periodic image of our quasiperiodic potetial, in a form of 
\begin{equation}\label{q-shro}
\frac{\partial^{2}}{\partial \xi^{2}}\psi(\xi,\eta)+ C\cdot(E-V(\xi,\eta)) \psi(\xi,\eta)=0, 
\end{equation}
where $E$ is the enregy and $C$ is a constant.


\section{Mathematical proof}
As it was said, quasiperiodic and periodic functions both might be expanded in a Fourier series. Using this fact, we will prove an equivalence between (\ref{q-shro}) and one-dimensional Schr\"{o}dinger equation with quasiperiodic potential.
\subsection{In one-dimensional space}
Acting like in Bloch theorem, substitute $\psi(x)=u(x)\cdot e^{iKx}$, where u(x) is quasiperiodic with quasiperiods the same as potential has, then we may obtain an equation for u(x):
\begin{multline*}\label{u-shro}
\frac{\partial^{2}}{\partial x^{2}}u(x)
+i2K
\frac{\partial}{\partial x}u(x)
+(E-V-K^{2})u(x)=0
\end{multline*}
Now expand $V(x)$
\begin{equation}\label{potential}
V(x)=\sum b_{lm} e^{i(l+\tau m)x}
\end{equation}
and $u(x)$ with coefficients $a_{lm}$. Using previously defined scalar product (\ref{scalar-multiply}), we finally get: 
\begin{multline} \label{1dim-coeff}
- a_{\alpha \beta}  \Big[\left(\alpha+\tau \beta+K\right) ^2-E\Big]
=\\=
\sum_{\begin{smallmatrix}l+m=\alpha\\n+k=\beta\end{smallmatrix}} 
a_{l m} b_{n k}
\end{multline}
\subsection{In two dimensions}
As previously, we firstly make a substitution
\begin{equation*}
\psi=u(x,y)\cdot e^{iKx+iNy}=u(\xi,\eta)\cdot e^{i\xi \frac{P}{2}}\cdot e^{i\eta \frac{K-\tau N}{2}}.
\end{equation*}
where $P=K+\tau N$ and u(x,y) is periodic. Substituting this to (\ref{q-shro}), we obtain
\begin{multline*}
\frac{\partial^{2}}{\partial \xi^{2}}u(\xi,\eta)
+iP\frac{\partial}{\partial \xi}u(x)
+C(E-V)
-\frac{P^{2}}{4} u(x)=0
\end{multline*}
Then, making orthogonalization procedure with standard scalar product in $L^2$, 
\begin{multline*} \label{2dim-coeff}
- a_{\alpha \beta}  \Big[\left(l+\tau m+P\right) ^2-4\cdot C \cdot E\Big]
=
4\cdot C \sum_{\begin{smallmatrix}l+m=\alpha\\n+k=\beta\end{smallmatrix}} 
a_{l m} b_{n k}
\end{multline*}
Comparing with (\ref{1dim-coeff}) we may conclude that C must be equal to $\frac{1}{4}$ and, finally, our equation takes the form
\begin{equation}\label{q-shro-full}
\frac{\partial^{2}}{\partial \xi^{2}}\psi(\xi,\eta)+ \frac{1}{4}(E-V(\xi,\eta)) \psi(\xi,\eta)=0
\end{equation}
and is totally equivalent to the Schr\"{o}dinger equation in one-dimensional space and quasiperiodic potential.

\section{Application: screened Coulomb potential}
I have applied this new technique to obtain the spectrum of non-interacting electrons located in quasiperiodic screened-Coulomb (Debye) potential.
\begin{equation*}\label{debay-pot-1d} 
V(x)=
\sum_{+\infty}^{+\infty} V_{1}\frac{e^{-\frac{x_{n}}{r_d}}}{x_{n}}
+V_{2}\frac{e^{-\frac{x'_{n}}{r_d}}}{x'_{n}}
\end{equation*}
where $x_{n}=|x-2\pi n|$, n is an integer. $x'_n= |\frac{x-2\pi \tau n}{\tau}|$, where $\tau=\frac{\sqrt 5+1}{2}$. 
After projecting, it becomes periodic in $[-\pi,\pi] \times [-\pi,\pi]$.\\

It was solved numerically in finite differences in a cell $[-\pi, \pi]  \times [-\pi, \pi]$.
For $V1=-0.5$, $V2=-0.7$ and $r_d=1$ no allowed negative energy levels were found. Spectrum for positive energies, just like it is in Schr\"{o}dinger operator with quasiperiodic potential \cite{simon} has no isolated points, and just like in previous works \cite{kitaev} it has a sophisticated Cantor-like structure.

\begin{center}
\centering\includegraphics[height=6in]{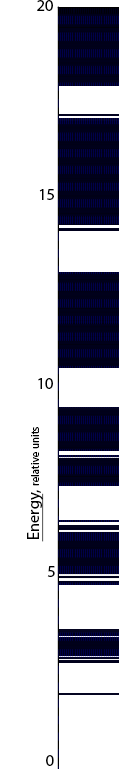}\\
\small \textit{Fig.3 Spectrum for a screened Coulomb potental}
\label{spectrum}
\end{center}
For every level we got the 2-dimensional images of wavefunctions. For described potential all of them were delocalized.  
\begin{center}
\centering\includegraphics[width=2in]{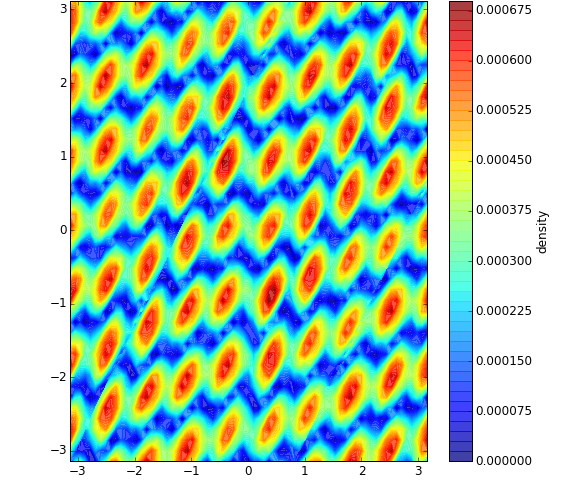}\\
\small \textit{ Fig.4  2-dimensional image of  wavefunction\\ with with E=18.59 and K=0.11}
\label{wave1}
\end{center}
It appeared that harmonics with higher frequencies are becoming more important with increasing an energy (Fig 4 and 5).
\begin{center}
\centering\includegraphics[width=2in]{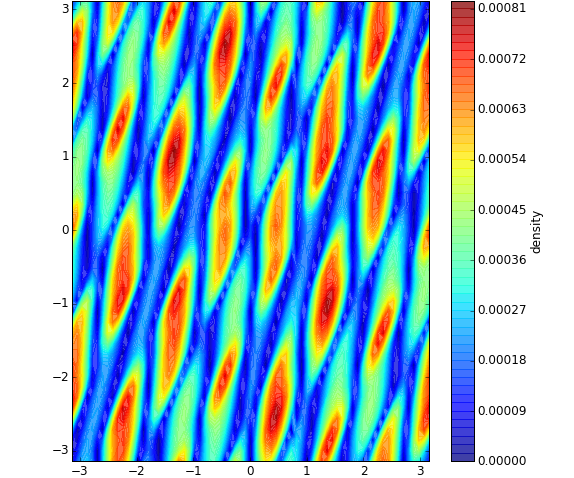}\label{wave2}\\
\small \textit{Fig.5  2-dimensional image of wavefunction\\ with E=10.55 and K=0.11}

\end{center}

\section{Generalization}
Using Fourier series, it is easy to generalize our equation to quasicrystals with higher dimensionality and arbitrary symmetries.\\ Firstly, let us expand our three dimensional function into series
\begin{equation*}
\psi=\sum a_{k_1 k_2 ... k_{n-1} k_n}e^{i(\hat{S}\cdot H)^T \cdot(x,y,z)^T}\cdot e^{iKx+iNy+iLz},
\end{equation*} where $\hat{S}$ is the transformation matrix performing the slicing opertaion \cite{plagiator}

\[ \hat{S}= \left( \begin{array}{cccccc}
\tau_{11} & 0 & 0 & \tau_{14} & ... & \tau_{1n}\\
0 & \tau_{22} & 0 & \tau_{24} & ... & \tau_{2n}\\
0 & 0 & \tau_{33} & \tau_{34} &... & \tau_{3n}\\
\end{array} \right)\] 
So, the most general form of kinetic part is exactly:
\begin{equation}\label{general-q-shro}
H_{kin}=\sum_{i,j}\frac{\partial^{2}}{\partial x_i \partial x_j}\cdot\Big(\tau_{1i}\tau_{1j}+\tau_{2i}\tau_{2j}+\tau_{3i}\tau_{3j}\Big)
\end{equation} 
This equation, suitable for any quasiperiodic potentials is the main result of the work.\\
Equation can also be applied to objects periodical in one or several dimensions. We just need to zero out corresponding elements in $\hat{S}$.
As an example, we may do it for an object that has 12-fold symmetry in x-y and periodic along z-axis.
In this specific case \cite{plagiator}, we will have $\hat{S}$ equal to
\[ \left( \begin{array}{ccccc}
1 & 0 & 0 & cos\frac{\pi}{6} & cos\frac{\pi}{3}\\
0 & 1 & 0 & sin\frac{\pi}{6} & sin\frac{\pi}{3}\\
0 & 0 & 1 & 0 & 0\\
\end{array} \right)\] 
Starting from here one may find the kinetic part of our Hamiltonian
\begin{multline*}\label{two-dimensional-H}
H_{kin}=\frac{\partial^{2}}{\partial x^{2}}+
\frac{\partial^{2}}{\partial y^{2}}+
\frac{\partial^{2}}{\partial z^{2}}+
\frac{\partial^{2}}{\partial \Theta^{2}}+
\frac{\partial^{2}}{\partial \Lambda^{2}}+\\+
2\frac{\partial}{\partial x}\cdot \Big(\cos\frac{\pi}{6}\cdot \frac{\partial}{\partial \Theta}+\cos\frac{\pi}{3}\cdot\frac{\partial}{\partial \Lambda}\Big)+\\+
2\frac{\partial}{\partial y}\cdot \Big(\sin\frac{\pi}{6}\cdot \frac{\partial}{\partial \Theta}+\sin\frac{\pi}{3}\cdot\frac{\partial}{\partial \Lambda}\Big)+\\+
2\frac{\partial}{\partial \Lambda} \frac{\partial}{\partial \Theta}\cdot\Big(\sin\frac{\pi}{6} \sin\frac{\pi}{3} +\cos\frac{\pi}{6} \cos\frac{\pi}{3}\Big)
\end{multline*}
 
\section{Conclusions}
This work overcame the problem of periodic description quasiperiodic wave-functions of non-interacting electrons. \\
Our method gives an ability easily get all the information about non-interacting electrons in arbitrary quasiperiodic potential.\\
Numerical results for one-dimensional quasiperiodic Debye-potential were obtained. Spectrum of this chain has self-similarity properties and likely has no isolated points. These properties agree with the general theorem \cite{simon} which indicates corectness of the method. \\
Since the generalization of the method is shown, results for real quasicrystalline materials can be obtained.\\
\section{Acknowledgments }
The author is very grateful to all his colleagues (especially to A.R. Oganov, Q. Fei , A.M. Komendantyan , Yu. Fomin, and  P. Dolgirev) for their interest and useful comments.
Work is funded by grant of Government of the Russian Federation No 14.A12.31.0003.


\end{multicols}

\end{document}